\documentclass[a4paper,USenglish,cleveref,autoref]{lipics-v2021}
\usepackage[utf8]{inputenc}
\usepackage[T1]{fontenc}
\usepackage{microtype}
\usepackage{amsfonts}
\usepackage{amssymb}
\usepackage{amsmath}
\usepackage{amsthm}
\usepackage{booktabs}
\usepackage{soul}

\newcommand{\LCE}{\mathrm{LCE}}

\newcommand{\bigO}{\mathcal{O}}

\DeclareMathOperator{\polylog}{polylog}
\newcommand{\height}[1]{\mathrm{height}(#1)}
\newcommand{\myexp}[1]{\mathrm{exp}(#1)}
\newcommand{\rhs}[1]{\mathrm{rhs}(#1)}
\newcommand{\sym}[1]{\mathrm{sym}(#1)}
\newcommand{\Decompose}{\mathsf{Decompose}}
\newcommand{\DecomposeWithRoots}{\mathsf{DecomposeWithRoots}}
\newcommand{\AddMerged}{\mathsf{AddMerged}}
\newcommand{\AddSymbol}{\mathsf{AddSymbol}}
\newcommand{\AddSubstring}{\mathsf{AddSubstring}}
\newcommand{\MergeEnclosed}{\mathsf{MergeEnclosed}}

\newcommand{\T}{T}

\newcommand{\dd}{\mathinner{.\,.}}
\newcommand{\eps}{\varepsilon}

\title{Fast and Space-Efficient Construction of AVL Grammars
  from the LZ77 Parsing}
\titlerunning{Fast and Space-Efficient Construction of AVL
  Grammars}

\author{Dominik Kempa}{Department of Computer Science,
  Johns Hopkins University, Baltimore, MD, USA}
  {kempa@cs.jhu.edu}{https://orcid.org/0000-0003-2286-7417}{}
\author{Ben Langmead}{Department of Computer Science,
  Johns Hopkins University, Baltimore, MD, USA}
  {langmea@cs.jhu.edu}{https://orcid.org/0000-0003-2437-1976}{}

\authorrunning{D. Kempa and B. Langmead}
\Copyright{Dominik Kempa and Ben Langmead}



\supplement{\url{https://github.com/dominikkempa/lz77-to-slp}}
\funding{DK and BL were supported by NIH HG011392 and NSF DBI-2029552 grants.}
\nolinenumbers

\EventEditors{}
\EventNoEds{2}
\EventLongTitle{}
\EventShortTitle{}
\EventAcronym{}
\EventYear{}
\EventDate{}
\EventLocation{}
\EventLogo{}
\SeriesVolume{}
\ArticleNo{1}

\begin{document}

\maketitle

\begin{abstract}
  Grammar compression is, next to Lempel-Ziv (LZ77) and run-length
  Burrows-Wheeler transform (RLBWT), one of the most flexible
  approaches to representing and processing highly compressible
  strings.  The main idea is to represent a text as a context-free
  grammar whose language is precisely the input string. This is called
  a straight-line grammar (SLG). An AVL grammar, proposed by Rytter
  [Theor. Comput. Sci., 2003] is a type of SLG that additionally
  satisfies the AVL-property: the heights of parse-trees for children
  of every nonterminal differ by at most one.  In contrast to other
  SLG constructions, AVL grammars can be constructed from the LZ77
  parsing in compressed time: $\bigO(z \log n)$ where $z$ is the size
  of the LZ77 parsing and $n$ is the length of the input text. Despite
  these advantages, AVL grammars are thought to be too large to be
  practical.

  We present a new technique for rapidly constructing a small AVL
  grammar from an LZ77 or LZ77-like parse.  Our algorithm produces
  grammars that are always at least five times smaller than those
  produced by the original algorithm, and never more than double the
  size of grammars produced by the practical Re-Pair compressor
  [Larsson and Moffat, Proc. IEEE, 2000].  Our algorithm also achieves
  low peak RAM usage.  By combining this algorithm with recent
  advances in approximating the LZ77 parsing, we show that our method
  has the potential to construct a run-length BWT from an LZ77 parse
  in about one third of the time and peak RAM required by other
  approaches.  Overall, we show that AVL grammars are surprisingly
  practical, opening the door to much faster construction of key
  compressed data structures.
\end{abstract}

\section{Introduction}\label{sec:intro}

The increase in the amount of highly compressible data that requires
efficient processing in the recent years, particularly in the area of
computational genomics~\cite{phoni, BoucherGKLMM19}, has caused a
spike of interest in dictionary compression. Its main idea is to
reduce the size of the representation of data by finding repetitions
in the input and encoding them as references to other
occurrences. Among the most popular methods are the Lempel-Ziv (LZ77)
compression~\cite{LZ77}, run-length Burrows-Wheeler transform
(RLBWT)~\cite{BWT, Gagie2020}, and grammar
compression~\cite{charikar}. Although in theory, LZ77 and RLBWT are
separated by at most a factor of $\bigO(\log^2 n)$ (where $n$ is the
length of the input text)~\cite{GNPlatin18, focs2020}, the gap in
practice is usually noticeable (as also confirmed by our experiments).
RLBWT is the largest of the three representations in practice, but is
also the most versatile, supporting powerful suffix array and suffix
tree queries~\cite{Gagie2020}. LZ77, on the other hand, is the
smallest, but its functionality includes only the easier LCE,
random-access, and pattern matching queries~\cite{DCC2015, ClaudeN11,
GagieGKNP14, GagieGKNP12, attractors}.  Grammar compression occupies
the middle ground between the two, supporting queries similar to
LZ77~\cite{Navarro2021b}. Navarro gives a comprehensive overview of
these and related representations in~\cite{Navarro2021b, Navarro2021}.

A major practical concern with these representations -- RLBWT in
particular -- is how to construct them efficiently.  Past efforts have
focused on engineering efficient \emph{general} algorithms for
constructing the BWT and LZ77~\cite{fgm2012, pSAscan, eSAIS,
kkp2014-dcc-lz}, but these are not applicable to the terabyte-scale
datasets routinely found, e.g., in modern
genomics~\cite{BoucherGKLMM19}.  Specialized algorithms for highly
repetitive datasets have only been investigated recently.  Boucher et
al.~\cite{BoucherGKLMM19} proposed a method for the efficient
construction of RLBWT using the concept of prefix-free parsing. The
same problem was approached by Policriti and Prezza, and Ohno et
al.~\cite{PolicritiP18, OhnoSTIS18}, using a different approach based
on the dynamic representation of RLBWT. These methods represent the
state of the art in the practical construction of RLBWT.

A different approach to the construction of RLBWT was recently
proposed in~\cite{focs2020}. The idea is to first compute the (exact
or approximate) LZ77 parsing for the text, and then convert this
representation into an RLBWT. Crucially, the LZ77 $\rightarrow$ RLBWT
conversion takes only $\bigO(z \polylog n)$ time, i.e. it runs not
only in the compressed space but also in \emph{compressed
time}.~\footnote{$\polylog n = \log^c n$ for any constant $c>0$.}
The computational bottleneck is therefore shifted to the easier
problem of computing or approximating the LZ77, which is the only step
taking $\Omega(n)$ time. Internally, this new pipeline consists of
three steps: text $\rightarrow$ (approximate) LZ77$\rightarrow$
grammar $\rightarrow$ RLBWT, unsurprisingly aligning with the gradual
increase in the size and complexity of these
representations. Kosolobov et al.~\cite{relz} recently proposed a fast
and space-efficient algorithm to approximate LZ77, called Re-LZ. The
second and third steps in the pipeline, from the LZ77 parse to the
RLBWT, have not been implemented. The only known algorithm to convert
LZ77 into a grammar in compressed time was proposed by
Rytter~\cite{Rytter03}, and is based on the special type of grammars
called \emph{AVL grammars}, whose distinguishing feature is that all
subtrees in the parse tree satisfy the AVL property: the tree-heights
for children of every nonterminal do not differ by more than one. The
algorithm is rather complex, and until now has been considered
impractical.

\subparagraph*{Our Contribution.}

Our main contribution is a series of practical improvements to the
basic variant of Rytter's algorithm, and a fast and space-efficient
implementation of this improved algorithm. Compared to the basic
variant, ours produces a grammar that is always five times smaller,
and crucially, the same holds for all intermediate grammars computed
during the algorithm, yielding very low peak RAM usage.  The resulting
grammar is also no more than twice of the smallest existing grammar
compressors such as Re-Pair~\cite{repair}.  We further demonstrate
that combining our new improved algorithm with Re-LZ opens up a new
path to the construction of RLBWT. Our preliminary experiments
indicate that at least a three-fold speedup and the same level of
reduction in the RAM usage is possible.

The key algorithmic idea in our variant is to delay the merging of
intermediate AVL grammars as much as possible to avoid creating
nonterminals that are then unused in the final grammar. We dub this
variant \emph{lazy AVL grammars}. We additionally incorporate
Karp-Rabin fingerprints~\cite{KR} to re-write parts of the grammar
on-the-fly and further reduce the grammar size. We describe two
distinct versions of this technique: greedy and optimal, and
demonstrate that both lead to reductions in the grammar size.  As a
side-result of independent interest, we describe a fast and
space-efficient data structure for the dynamic predecessor problem, in
which the inserted key is always larger than all other elements
currently in the set.

\section{Preliminaries}\label{sec:prelim}

\subparagraph*{Strings.}

For any string $S$, we write $S[i\dd j]$, where $1 \leq i,j \leq |S|$,
to denote a substring of $S$. If $i>j$, we assume $S[i\dd j]$ to be
the empty string $\eps$. By $[i \dd j)$ we denote $[i \dd j - 1]$.
Throughout the paper, we consider a string (text) $\T[1\dd n]$ of $n
\geq 1$ symbols from an integer alphabet $\Sigma = [0 \dd \sigma)$.
By $\LCE(i, i')$ we denote the length of the longest common prefix of
suffixes $\T[i \dd n]$ and $\T[i' \dd n]$.

\vspace{-1ex}
\subparagraph*{Karp-Rabin Fingerprints.}

Let $q$ be a prime number and let $r \in [0 \dd q - 1]$ be chosen
uniformly at random. The \emph{Karp-Rabin fingerprint} of a string $S$
is defined as
\[
  \Phi(S) = \sum_{i = 1}^{|S|} S[i]\cdot r^{|S| - i} \bmod q\ .
\]
Clearly, if $\T[i \dd i + \ell) = \T[j \dd j + \ell)$ then $\Phi(\T[i
\dd i + \ell)) = \Phi(\T[j \dd j + \ell))$. On the other hand, if
$\T[i \dd i + \ell) \neq \T[j \dd j + \ell)$ then $\Phi(\T[i \dd i +
\ell)) \neq \Phi(\T[j \dd j + \ell))$ with probability at least $1 -
\ell/q$~\cite{DGMP92}. In our algorithm we are comparing only
substrings of $\T$ of equal length. Thus, the number of different
possible substring comparisons is less than~$n^3$, and hence for any
positive constant $c$, we can set $q$ to be a prime larger than $n^{c
+ 4}$ (but still small enough to fit in $\bigO(1)$ words) to make
the fingerprint function perfect with probability at least $1 -
n^{-c}$.

\vspace{-1ex}
\subparagraph*{LZ77 Compression.}

We say that a substring $\T[i\dd i+\ell)$ is a \emph{previous factor}
if it has an earlier occurrence in $\T$, i.e., there exists a position
$p \in [1 \dd i)$ such that $\LCE(i,p)\ge \ell$. An \emph{LZ77-like
factorization} of $\T$ is a factorization $\T = F_1\cdots F_f$ into
non-empty \emph{phrases} such that every phrase $F_j$ with $|F_j|>1$
is a previous factor. Every phrase $F_j=\T[i\dd i+\ell)$ is encoded as
a pair $(p,\ell)$, where $p\in [1\dd i)$ is such that $\LCE(i,p)\ge
\ell$. If there are multiple choices for such $p$, we choose one
arbitrarily. The occurrence $\T[p \dd p + \ell)$ is called the
\emph{source} of the phrase $F_j$. If $F_j = T[i]$ is not a previous
factor, it is encoded as a pair $(T[i],0)$.

The LZ77 factorization~\cite{LZ77} (or the LZ77 parsing) of a string
$\T$ is an LZ77-like factorization constructed by greedily parsing
$\T$ from left to right into longest possible phrases.  More
precisely, the $j$th phrase $F_j$ is the longest previous factor
starting at position $ i = 1+|F_1\cdots F_{j-1}|$. If no previous
factor starts there, then $F_j = \T[i]$.  We denote the number of
phrases in the LZ77 parsing by $z$.  For example, the text
$\texttt{bbabaababababaababa}$ has LZ77 parsing $\texttt{b}\cdot
\texttt{b}\cdot \texttt{a}\cdot \texttt{ba} \cdot \texttt{aba}\cdot
\texttt{bababa} \cdot \texttt{ababa}$ with $z=7$ phrases, and is
encoded as a sequence $ (\texttt{b},0), (1,1), (\texttt{a},0), (2,2),
(3,3), (7,6), (10,5).  $

\vspace{-1ex}
\subparagraph*{Grammar Compression.}

A context-free grammar is a tuple $G = (N, \Sigma, R, S)$, where $N$
is a finite set of \emph{nonterminals}, $\Sigma$ is a finite set of
\emph{terminals}, and $R \subseteq N \times (N \cup \Sigma)^*$ is a
set of \emph{rules}. We assume $N \cap \Sigma = \emptyset$ and $S \in
N$. The nonterminal $S$ is called the \emph{starting symbol}.  If $(A,
\gamma) \in R$ then we write $A \rightarrow \gamma$.  The
\emph{language} of $G$ is set $L(G) \subseteq \Sigma^*$ obtained by
starting with $S$ and repeatedly replacing nonterminals with their
expansions, according to $R$.

A grammar $G = (N, \Sigma, R, S)$ is called a \emph{straight-line
grammar} (SLG) if for any $A \in N$ there is exactly one production
with $A$ of the left side, and all nonterminals can be ordered $A_1,
\ldots, A_{|N|}$ such that $S = A_1$ and if $A_i \rightarrow \gamma$
then $\gamma \in (\Sigma \cup \{A_{i+1} \cup \ldots \cup
A_{|N|}\})^*$, i.e., the graph of grammar rules is acyclic. The unique
$\gamma$ such that $A \rightarrow \gamma$ is called the
\emph{definition} of $A$ and is denoted $\rhs{A}$.  In any SLG, for
any $u \in (N \cup \Sigma)^*$ there exists exactly one $w \in
\Sigma^*$ that can be obtained from $u$. We call such $w$ the
\emph{expansion} of $u$, and denote $\myexp{u}$.  We define the
\emph{parse tree} of $A \in N \cup \Sigma$ as a rooted ordered tree
$\mathcal{T}(A)$, where each node $v$ is associated to a symbol
$\sym{v} \in N \cup \Sigma$. The root of $\mathcal{T}(A)$ is a node
$v$ such that $\sym{v} = A$. If $A \in \Sigma$ then $v$ has no
children. If $A \in N$ and $\rhs{A} = B_1 \cdots B_k$, then $v$ has
$k$ children and the subtree rooted at the $i$th child is a copy of
$\mathcal{T}(B_i)$. The parse tree $\mathcal{T}(G)$ is defined as
$\mathcal{T}(S)$. The \emph{height} of any $A \in N$ is defined as the
height of $\mathcal{T}(A)$, and denoted $\height{A}$.

The idea of \emph{grammar compression} is, given a text $T$, to
compute a small SLG $G$ such that $L(G) = T$. The \emph{size} of the
grammar is measured by the total length of all definitions, and
denoted $|G| := \sum_{A \in N}|\rhs{A}|$.  Clearly, it is easy to
encode any $G$ in $\bigO(|G|)$ space: pick an ordering of nonterminals
and write down the definitions of all variables with nonterminals
replaced by their number in the order.

\section{AVL Grammars and the Basic Algorithm}\label{sec:basic}

An SLG $G = (N, \Sigma, R, S)$ is said to be in \emph{Chomsky normal
form}, if for every $A \in N$, it holds $\rhs{A} \in \Sigma$ or
$\rhs{A} = XY$, where $X, Y \in N$.  An SLG in Chomsky normal form is
called a \emph{straight-line program} (SLP). Rytter~\cite{Rytter03}
defines an \emph{AVL grammar} as an SLP $G = (N, \Sigma, R, S)$ that
satisfies the following extra condition: for every $A \in N$ such that
$\rhs{A} = XY$, it holds $|\height{X}| - |\height{Y}| \leq 1$. This
condition guarantees that for every $A \in N$ (in particular for $S
\in N$), it holds $\height{A} = \bigO(\log
|\myexp{A}|)$~\cite[Lemma~1]{Rytter03}.

The main result presented in~\cite{Rytter03} is an algorithm that
given a non-self-referential version (in which $\T[i \dd i + \ell)$ is
a previous factor only if there exists $p \in [1 \dd i - \ell]$ such
that $\LCE(i, p) \geq \ell$) of some LZ77-like parsing for a text $\T$
of length $n$ consisting of $f$ phrases, computes in $\bigO(f \log n)$
time an AVL grammar $G$ generating $\T$ and satisfying $|G| = \bigO(f
\log n)$. Rytter's construction was extended to allow self-references
in~\cite[Theorem~6.1]{focs2020}. Our implementation of the basic as
well as improved Rytter's algorithm works for the self-referential
variant, but for simplicity here we only describe the algorithm for
the non-self-referential variant.

The algorithm in~\cite{Rytter03} works in $f$ steps. It maintains the
dynamically changing AVL grammar $G$ such that after the $k$th step is
complete, there exists a nonterminal $P_k$ in $G$ such that
$\myexp{P_k} = F_1 \cdots F_k$, where $\T = F_1 \cdots F_f$ is the
input LZ77-like factorization of the input. This implies that at end
there exist a nonterminal expanding to $\T$. The algorithm does not
delete any nonterminals between steps. At the end, it may perform an
optional pruning of the grammar to remove the nonterminals not present
in the parse tree $\mathcal{T}(P_f)$. This reduces the grammar size
but not the peak memory usage of the algorithm.

The algorithm uses the following three procedures, each of which adds
a nonterminal $A$ with a desired expansion $\myexp{A}$ to the grammar
$G$, along with a bounded number of extra nonterminals:

\begin{enumerate}
\item $\AddSymbol(c)$: Given $c \in \Sigma$, add a nonterminal $A$
  with $\rhs{A} = c$ to the grammar $G$.
\item $\AddMerged(X, Y)$: Given the identifiers of nonterminals $X, Y$
  existing in $G$, add a nonterminal $A$ to $G$ that satisfies
  $\myexp{A} = \myexp{X}\myexp{Y}$.  The difficulty of this operation
  is ensuring that the updated $G$ remains an AVL grammar. Simply
  setting $\rhs{A} = XY$ would violate the AVL condition in most
  cases. Instead, the algorithm performs the procedure similar to the
  concatenation of AVL trees~\cite[p.~474]{knuth}, taking $\bigO(1 +
  |\height{X} - \height{Y}|)$ time, and introducing $\bigO(|\height{X}
  - \height{Y}|)$ extra nonterminals.
\item $\AddSubstring(A, i, j)$: Given the identifier of a nonterminal
  $A$ existing in $G$, and two positions satisfying $1 \leq i \leq j
  \leq |\myexp{A}|$, add a nonterminal $B$ to $G$ that satisfies
  $\myexp{B} = \myexp{A}[i \dd j]$. To explain how this is achieved,
  we define the auxiliary procedure $\Decompose(A, i, j)$ that given
  the same parameters as above, returns the sequence $B_1, \ldots,
  B_q$ of nonterminals satisfying $\myexp{A}[i \dd j] = \myexp{B_1}
  \cdots \myexp{B_q}$.  The nonterminals $B_i$ are found by performing
  two root-to-leaf traversals in the parse tree $\mathcal{T}(A)$.
  This takes $\bigO(1 + \height{A}) = \bigO(1 + \log |\myexp{A}|)$
  time and ensures $q = \bigO(\log |\myexp{A}|)$. It is easy to see
  that given $B_1, \ldots, B_q$, we can now obtain $B$ in $\bigO(1 +
  \log^2 |\myexp{A}|)$ time using $\AddMerged$.  In~\cite{Rytter03},
  it was however shown that if we always choose the shortest
  nonterminal to merge with its neighbor, the total runtime is
  $\bigO(1 + \log |\myexp{A}|)$ and only $\bigO(\log |\myexp{A}|)$
  extra nonterminals are introduced.
\end{enumerate}

Using the above three procedures, the algorithm in~\cite{Rytter03}
works as follows. Suppose we have already processed the leftmost $k-1$
phrases.  The step begins by creating a nonterminal $A_k$ satisfying
$\myexp{A_k} = F_k$. If $|F_k| = 1$, it uses the procedure
$\AddSymbol(F_k)$. Otherwise, $A_k$ is obtained as the output of
$\AddSubstring(P_{k - 1}, p, p + \ell-1)$, where $\ell = |F_k|$ and
$\T[p \dd p + \ell)$ is the source of phrase $F_k$. Finally, $P_k$ is
obtained as the output of $\AddMerged(P_{k-1}, A_k)$. A single
iteration thus takes $\bigO(1 + \log |F_1 \cdots F_{k-1}|) =
\bigO(\log n)$ time and adds $\bigO(\log n)$ extra nonterminals, for
total of $\bigO(f \log n)$ nonterminals over all steps.

\section{Optimized Algorithm}\label{sec:optimized-alg}

\subparagraph*{Lazy Merging.}

We start by observing that the main reason responsible for the large
final grammar produced by the algorithm in \cref{sec:basic} is the
requirement that at the end of each step $k \in [1 \dd f]$, there
exist a nonterminal $P_k$ satisfying $\myexp{P_k} = F_1 \cdots
F_k$. We relax this requirement, and instead require only that at the
end of step $k$, there exists a sequence of nonterminals $R_1, \ldots,
R_m$ such that $\myexp{R_1} \cdots \myexp{R_m} = F_1 \cdots F_k$. The
algorithm explicitly maintains these nonterminals as a sequence of
pairs $(\ell_1, R_1), \ldots, (\ell_m, R_m)$, where $\ell_j =
\sum_{i=1}^{j}|\myexp{R_i}|$. The modified algorithm uses the
following new procedures:

\begin{enumerate}
\item $\MergeEnclosed(i, j)$: Given two positions satisfying $1 \leq i
  \leq j \leq |F_1 \cdots F_{k-1}|$, add to $G$ a nonterminal $R$
  satisfying $\myexp{R} = \myexp{R_x} \cdots \myexp{R_{y}}$, where $x
  = \min\{t \in [1 \dd m] : \ell_{t-1} \geq i - 1\}$ and $y = \max\{t
  \in [1 \dd m] : \ell_{t} \leq j\}$.  The positions $x$ and $y$ are
  found using a binary search. The pairs of the sequence $(\ell_1,
  R_1), \ldots, (\ell_m, R_m)$ at positions between $x$ and $y$ are
  removed and replaced with a pair ($\ell_y, R)$.  In other words,
  this procedure merges all the nonterminals from the current root
  sequence whose expansion is entirely inside the given interval $[i
  \dd j]$.  Merging of $R_x, \ldots, R_y$ is performed pairwise,
  using the $\AddMerged$ procedure. Note, however, that there is no
  dependence between heights of the adjacent nonterminals (in
  particular, they do not form a bitonic sequence, like in the
  algorithm in \cref{sec:basic}), and moreover, their number
  $\widehat{m} := y - x + 1$ is not bounded. To minimize the number of
  newly introduced nonterminals, we thus employ a greedy heuristic,
  that always chooses the nonterminal with the smallest height among
  the remaining elements, and merges it with the shorter neighbor. We
  use a list to keep the pointers between neighbors and a binary heap
  to maintain heights. The merging thus runs in $\bigO(\widehat{m}
  \log n)$ time.
\item $\DecomposeWithRoots(i, j)$: Given two positions satisfying $1
  \leq i \leq j \leq |F_1 \cdots F_{k-1}|$, this procedure returns a
  sequence of nonterminals $A_1, \ldots, A_q$ satisfying $(F_1 \cdots
  F_{k-1})[i \dd j] = \myexp{A_1} \cdots \myexp{A_q}$. First, it
  computes positions $x$ and $y$, as defined in the description of
  $\MergeEnclosed$ above. It then returns the result of
  $\Decompose(R_{x-1}, i - \ell_{x-2}, \ell_{x-1} - \ell_{x-2})$,
  followed by $R_x, \ldots, R_y$, followed by the result of
  $\Decompose(R_{y+1}, 1, j - \ell_{y})$ (appropriately handling the
  boundary cases, which for clarity we ignore here). In other words,
  this procedure finds the sequence of nonterminals that uses as many
  roots from the sequence $R_1, \ldots, R_m$, as possible, and runs
  the standard $\Decompose$ for the boundary roots. Letting
  $\widehat{m} = y - x + 1$, it runs in $\bigO(\widehat{m} + \log n)$
  time.
\end{enumerate}

Using the above additional procedures, our algorithm works as
follows. Suppose that we have already processed the first $k - 1$
phrases. The step begins by computing the sequence of nonterminals
$A_1, \ldots, A_q$ satisfying $\myexp{A_1} \cdots \myexp{A_q} =
F_k$. If $|F_k| = 1$, we proceed as in \cref{sec:basic}. Otherwise, we
first call $\MergeEnclosed(p, p + \ell - 1)$, where $\ell = |F_k|$ and
$\T[p \dd p + \ell)$ is the source of $F_k$. The sequence $A_1,
\ldots, A_q$ is then obtained as a result of $\DecomposeWithRoots(p, p
+ \ell - 1)$. Finally, $A_1, \ldots, A_q$, is appended to the roots
sequence.

The above algorithm runs in $\bigO(f \log^2 n)$ time. To see this,
note that first calling $\MergeEnclosed$ ensures that the output size
of $\DecomposeWithRoots$ is $\bigO(\log n)$.  Thus, each step appends
only $\bigO(\log n)$ nonterminals to the roots sequence.  The total
time spend in $\MergeEnclosed$ is thus bounded by $\bigO(f \log^2 n)$,
dominating the time complexity.

\vspace{-1ex}
\subparagraph*{Utilizing Karp-Rabin Fingerprints.}

Our second technique is designed to detect the situation in which the
algorithm adds a nonterminal $A$ to $G$, when there already exists
some $B \in N$ such that $\myexp{A} = \myexp{B}$. For any $u \in (N
\cup \Sigma)^{*}$, we define $\Phi(u) = \Phi(\myexp{u})$.  Let us
assume that there are no collisions between fingerprints.  Suppose
that given a nonterminal $A \in N$, we can quickly compute $\Phi(A)$,
and that given some $x \geq 0$, we can check, if there exists $B \in
N$ such that $\Phi(B) = x$. There are two places in the above
algorithm (using lazy merging) where we utilize this to reduce the
number of nonterminals:

\begin{enumerate}
\item Whenever during the greedy merge in $\MergeEnclosed$, we are
  about to call $\AddMerged$ for the pair of adjacent nonterminals $X$
  and $Y$, we instead first compute the fingerprint $x = \Phi(XY)$ of
  their concatenation, and if there already exists $A \in N$ such that
  $\Phi(A) = x$, we use $A$ instead, avoiding the call to $\AddMerged$
  and the creation of extra nonterminals.
\item Before appending the nonterminals $A_1, \ldots, A_q$ to the
  roots sequence at the end of the step, we check if there exists an
  equivalent but shorter sequence $B_1, \ldots, B_{q'}$, i.e., such
  that $\myexp{A_1} \cdots \myexp{A_q} = \myexp{B_1} \cdots
  \myexp{B_{q'}}$ and $q' < q$. We utilize that $q = \bigO(\log n)$,
  and run a quadratic algorithm (based on dynamic programming) to find
  the optimal (shortest) equivalent sequence. We then use that
  equivalent sequence in place of $A_1, \ldots, A_q$.
\end{enumerate}

Observe that the above techniques cannot be applied during
$\AddMerged$, as the equivalent nonterminal could have a notably
shorter/taller parse tree, violating the AVL property. Note also that
the algorithm remains correct even when checking if there exists $A
\in N$ having $\Phi(A) = x$ returns false negatives. Our
implementation uses a parameter $p \in [0 \dd 1]$, which is the
probability of storing the mapping from $\Phi(A)$ to $A$, when adding
the nonterminal $A$. The value of $p$ is one of the main parameters
controlling the time-space trade-off of the algorithm, as well as the
size of the final grammar. The mapping of fingerprints to nonterminals
is implemented as a hash table.

\section{Implementation Details}\label{sec:implementation}

\subparagraph*{Storing Sequences.}

Our implementation stores many sequences, where the insertion only
happens at the end (e.g., the sequence of nonterminals, which are
never deleted in the algorithm). A standard approach to this is to use
a \emph{dynamic array}, which is a plain array that doubles its
capacity, once it gets full. Such implementation achieves an amortized
$\bigO(1)$ insertion time, but suffers from a high peak RAM usage. On
all systems we tried, the reallocation call that extends the array is
not in-place. Since the peak RAM usage is critical in our
implementation, we thus implemented our own dynamic array, that
instead of a single allocated block, keeps a larger number of blocks
(we use 32). This significantly reduces the peak RAM usage.  We found
the slowdown in the access-time to be negligible.

\subparagraph*{Implementation of the Roots Sequence.}

The roots sequence $(\ell_1, R_1), \ldots, (\ell_m, R_m)$ undergoes
predecessor queries, deletions (at arbitrary positions), and
insertions (only at the end). Rather than using an off-the-shelf
dynamic predecessor data structure (such as balanced BST), we exploit
as follows the fact that insertions happen only at the end.

All roots are stored as a static sequence that only undergoes
insertions at the end (using the space efficient dynamic array
implementation described above). Deleted elements are marked as
deleted, but remain physically in the array. The predecessor query is
implemented using a binary search, with skipping of the elements
marked as deleted. To ensure that the predecessor queries are
efficient, we keep a counter of accesses to the deleted elements.
Once it reaches the current array size, we run the ``garbage
collector'' that scans the whole array left-to-right, and removes all
the elements marked as deleted, eliminating all gaps. This way, the
predecessor query is still efficient, except the complexity becomes
amortized.

\subparagraph*{Computing $\Phi(A)$ and $|\myexp{A}|$ for $A \in N$.}

During the algorithm, we often need to query the value of $\Phi(A)$
for some nonterminal $A \in N$. In our implementation we utilize
64-bit fingerprints, and hence storing the value $\Phi(A)$ for every
nonterminal is expensive. We thus only store $\Phi(A)$ for $A \in N$
satisfying $|\myexp{A}| \geq 255$. The number of such elements in $N$
is relatively small. To compute $\Phi(A)$ for any other $A \in N$, we
first obtain $\myexp{A}$, and then compute $\Phi(A)$ from
scratch. This operation is one of the most expensive in our algorithm,
and hence whenever possible we avoid doing repeated $\Phi$ queries.

As for the values $|\myexp{A}|$, we observe that in most cases, it
fits in a single byte. Thus, we designate only a single byte, and
whenever $|\myexp{A}| \geq 255$, we lookup $|\myexp{A}|$ in an array
ordered by the number of nonterminal, with access implemented using
binary search.

\section{Experimental Results}\label{sec:experiments}

\subparagraph*{Algorithms.}

We performed experiments using the following algorithms:
\begin{itemize}
\item \ul{Basic-AVLG}, our implementation of the algorithm to convert
  an LZ-like parsing to an AVL grammar proposed by
  Rytter~\cite{Rytter03}, and outlined in \cref{sec:basic}.
  Self-overlapping phrases are handled as in the full version
  of~\cite[Theorem~6.1]{focs2020}. The implementation uses
  space-efficient dynamic arrays described in
  \cref{sec:implementation}. This implementation is our~baseline.
\item \ul{Lazy-AVLG}, our implementation of the improved version of
  Basic-AVLG, utilizing lazy merging and Karp-Rabin fingerprints, as
  described in \cref{sec:optimized-alg}. This is the main contribution
  of our paper.  In some of the experiments below, we consider the
  algorithm with the different probability $p$ of sampling the
  Karp-Rabin hash of a nonterminal, but our default value (as
  discussed below) is $p = 0.125$.  Our implementation (including also
  Basic-AVLG) is available at
  \url{https://github.com/dominikkempa/lz77-to-slp}.
\item \ul{Big-BWT}, a semi-external algorithm constructing the RLBWT
  from the input text in $\Omega(n)$ time, proposed by Boucher et
  al.~\cite{BoucherGKLMM19}. As shown in~\cite{BoucherGKLMM19}, if the
  input text is highly compressible, the working space of Big-BWT is
  sublinear in the text length. We use the implementation from
  \url{https://gitlab.com/manzai/Big-BWT}.
\item \ul{Re-LZ}, an external-memory algorithm due to Kosolobov et
  al.~\cite{relz} that given a text on disk, constructs its LZ-like
  parsing in $\Omega(n)$ time~\cite{relz}. The algorithm is faster
  than the currently best algorithms to compute the LZ77 parsing.  In
  practice, the ratio $f/z$ between the size $f$ of the resulting
  parsing and the size $z$ of the LZ77 parsing usually does not exceed
  1.5. Its working space is fully tunable and can be specified
  arbitrarily. We use the implementation from
  \url{https://gitlab.com/dvalenzu/ReLZ}.
\item \ul{Re-Pair}, an $\bigO(n)$-time algorithm to construct an SLG
  from the text, proposed by Larsson and
  Moffat~\cite{repair}. Although no upper-bound is known on its output
  size, Re-Pair produces grammars that in practice are smaller than
  any other grammar-compression method~\cite{repair}. Its main
  drawback is that most implementations need $\Theta(n)$
  space~\cite{FuruyaTNIBK19}, and hence are not applicable on massive
  datasets. The only implementation using $o(n)$ space
  is~\cite{KopplIFTSG21}, but it is not practical. There is also work
  on running Re-Pair on the compressed input~\cite{SakaiOGTIS19}, but
  since it already requires the text as a grammar, it is not
  applicable in our case.  In our experiments we therefore use Re-Pair
  only as a baseline for the achievable grammar size. We use our own
  implementation, verified to produce grammars comparable to other
  implementations. We note that there exists recent work on optimizing
  Re-Pair by utilizing maximal repeats~\cite{FuruyaTNIBK19}. The
  decrease in the grammar size, however, requires a potentially more
  expensive encoding that includes the length of the expansion for
  every nonterminal. For simplicity, we therefore use the basic
  version of Re-Pair.
\end{itemize}

\begin{table*}
  \centering
  \caption{Statistics of files used in the experiments, with $n$
    denoting text length, $\sigma$ denoting alphabet size, $r$
    denoting the number of runs in the BWT, and $z$ denoting the
    number of phrases in the LZ77 parsing. For convenience, we also
    show the average BWT run length $n/r$ and the average LZ77 phrase
    length $n/z$. Each of the symbols in the input texts is encoded
    using a single byte.}
  \label{tab:files}
  \setlength{\tabcolsep}{6pt}
  \renewcommand{\arraystretch}{0.85}
  \begin{tabular}{lrrrrrr}
    \toprule
    File name        &                $n$ &   $\sigma$ &           $r$ &   $n/r$ &           $z$ &     $n/z$ \\
    \midrule
    cere             &      461\,286\,644 &          5 &  11\,574\,640 &   39.85 &   1\,700\,630 &    271.24 \\
    coreutils        &      205\,281\,778 &        236 &   4\,684\,459 &   43.82 &   1\,446\,468 &    141.91 \\
    einstein.en.txt  &      467\,626\,544 &        139 &      290\,238 & 1611.18 &       89\,467 & 5\,226.80 \\
    influenza        &      154\,808\,555 &         15 &   3\,022\,821 &   51.21 &      769\,286 &    201.23 \\
    dna.001.1        &      104\,857\,600 &          5 &   1\,716\,807 &   61.07 &      308\,355 &    340.05 \\
    english.001.2    &      104\,857\,600 &        106 &   1\,449\,518 &   72.33 &      335\,815 &    312.24 \\
    proteins.001.1   &      104\,857\,600 &         21 &   1\,278\,200 &   82.03 &      355\,268 &    295.15 \\
    sources.001.2    &      104\,857\,600 &         98 &   1\,213\,427 &   86.41 &      294\,994 &    355.45 \\
    chr19.1000       &  59\,125\,116\,167 &          5 &  45\,927\,063 & 1287.37 &   7\,423\,960 &   7964.09 \\
    kernel           & 137\,438\,953\,472 &        229 & 129\,506\,377 & 1061.25 &  30\,222\,602 &   4547.55 \\
    \bottomrule
  \end{tabular}
\end{table*}

All implementations are in C++ and are largely sequential, allowing
for a constant number of additional threads used for asynchronous I/O.

We also considered Online-RLBWT, an algorithm proposed by Ohno at
al.~\cite{OhnoSTIS18}, that given a text in a right-to-left streaming
fashion, construct its run-length compressed BWT (RLBWT) in $\bigO(n
\log r)$ time and using only $\bigO(r \log n)$ bits of working space
(the implementation is available from:
\url{https://github.com/itomomoti/OnlineRlbwt}). In the preliminary
experiment we determined that while using only about a third of the
memory of Big-BWT (on the 16\,GiB prefix of the kernel testfile), the
algorithm was about 10x slower than Big-BWT.

\subparagraph*{Experimental Platform and Datasets.}

We performed experiments on a machine equipped with two twelve-core
2.2\,GHz Intel Xeon E5-2650v4 CPUs with 30\,MiB L3 cache and 512\,GiB
of RAM. The machine used distributed storage achieving an I/O rate
>220\,MiB/s (read/write).

The OS was Linux (CentOS 7.7, 64bit) running kernel 3.10.0.  All
programs were compiled using {\tt g++} version 4.8.5 with {\tt-O3}
{\tt-DNDEBUG} {\tt-march=native} options. All reported runtimes are
wallclock (real) times. The machine had no other significant CPU tasks
running. To measure the peak RAM usage of the programs we used the
{\tt /usr/bin/time -v} command.

The statistics of testfiles used in our experiments are shown in
\cref{tab:files}. Shorter version of files used in the scalability
experiments are prefixes of full files. We used the files from the
Pizza \& Chili repetitive corpus available at
\url{http://pizzachili.dcc.uchile.cl/repcorpus.html}.  We chose a
sample of 8 real and pseudo-real files. Since all files are relatively
small (less than 512\,MiB), we additionally include 2 large repetitive
files that we assembled ourselves:
\begin{itemize}
\item \ul{chr19.1000}, a concatenation of 1000 versions of Human
  chromosome 19. The sequences were obtained from the 1000 Genomes
  Project~\cite{1000k}. One copy consists of $\sim$58$\times 10^6$
  symbols.
\item \ul{kernel}, a concatenation of $\sim$10.7 million source files
  from over 300 versions of the Linux kernel (see
  \url{http://www.kernel.org/}).
\end{itemize}

\subparagraph*{Karp-Rabin Sampling Rate.}

\begin{figure}[t!]
  \centering
  \minipage{0.5\textwidth}
    \includegraphics[trim = 3mm 3mm 3mm 0mm, width=\linewidth]
      {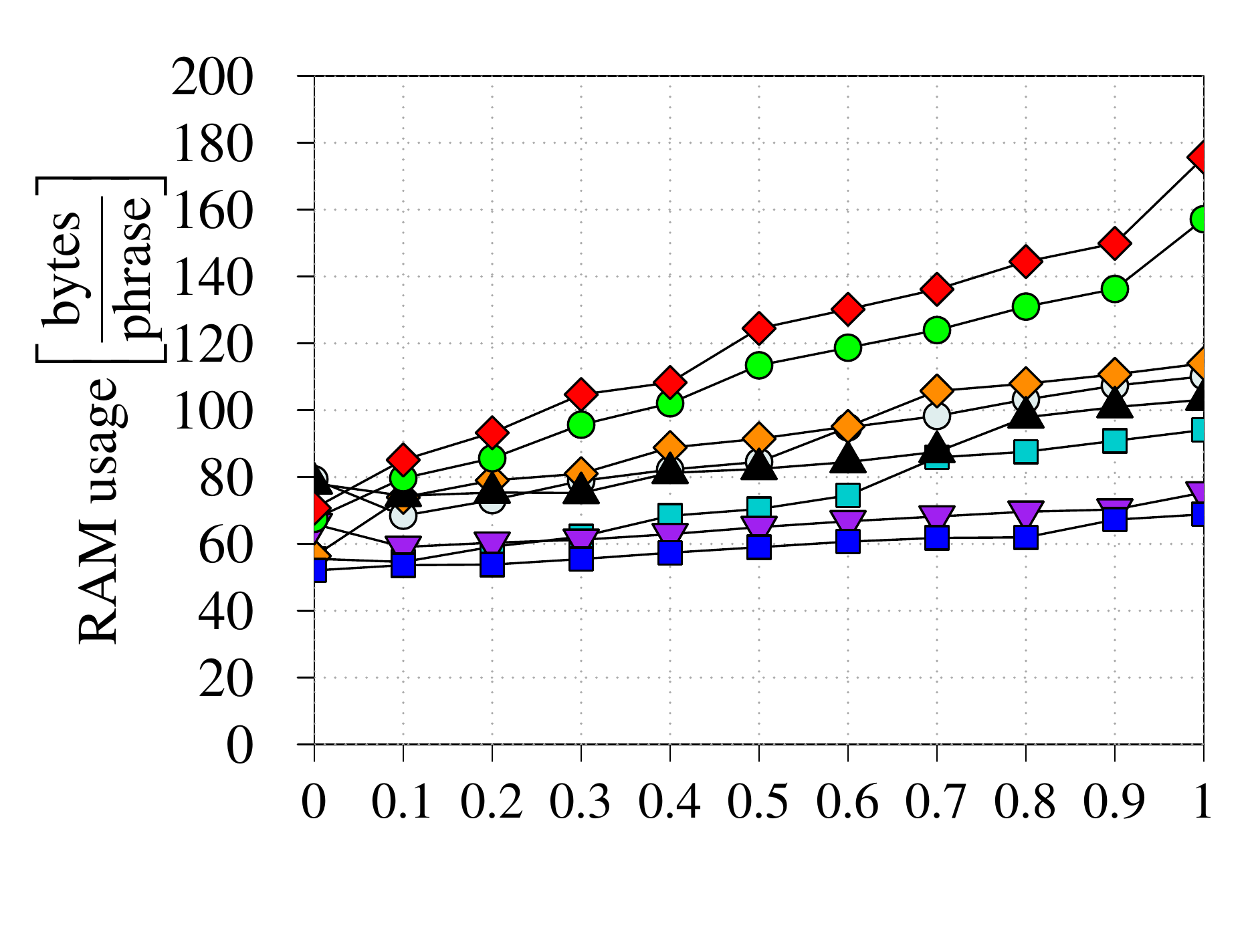}
  \endminipage
  \minipage{0.5\textwidth}
    \includegraphics[trim = 6mm 3mm 0mm 0mm, width=\linewidth]
      {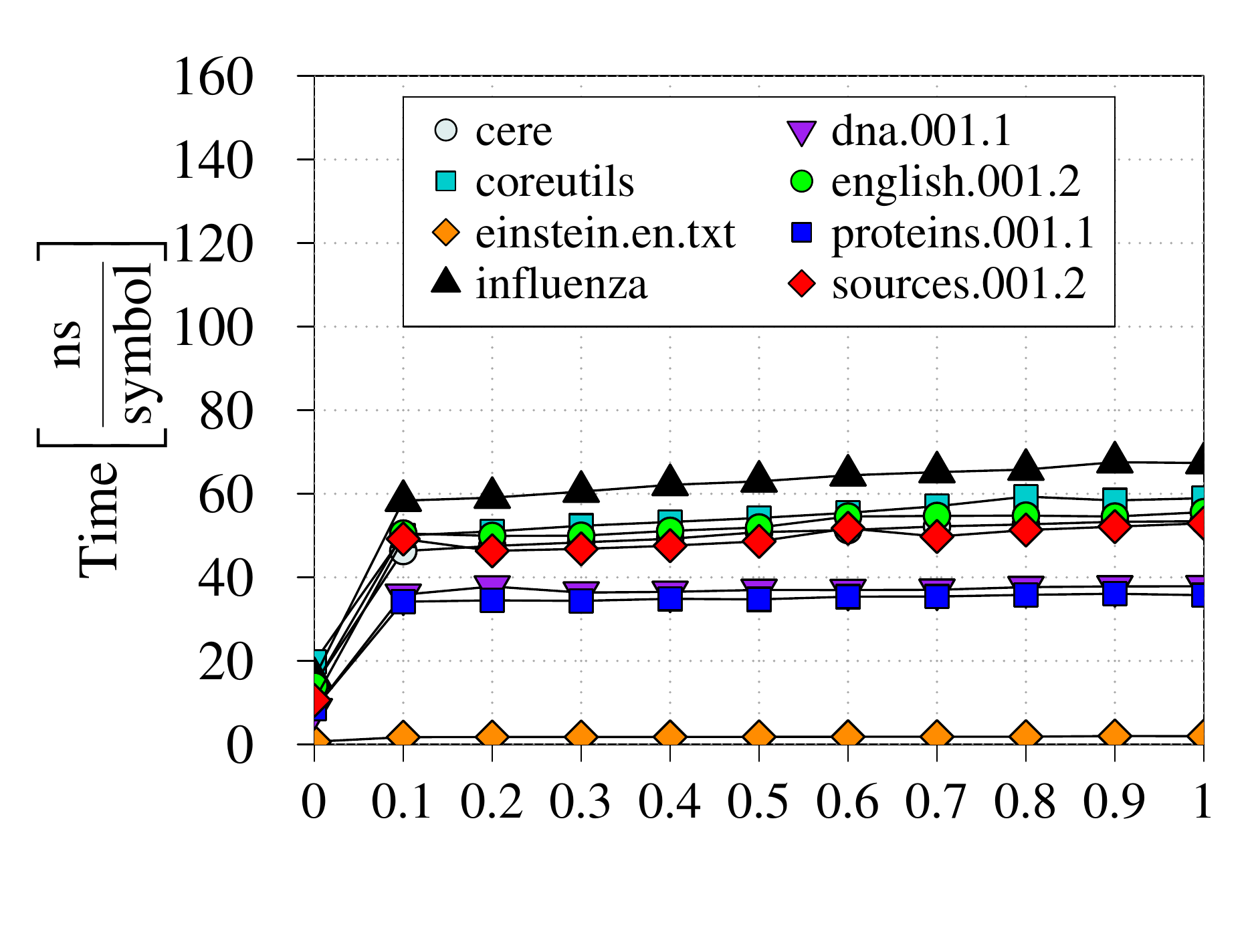}
  \endminipage
  \newline\vspace{-0.6cm}

  \minipage{0.5\textwidth}
    \includegraphics[trim = 3mm 3mm 3mm 0mm, width=\linewidth]
      {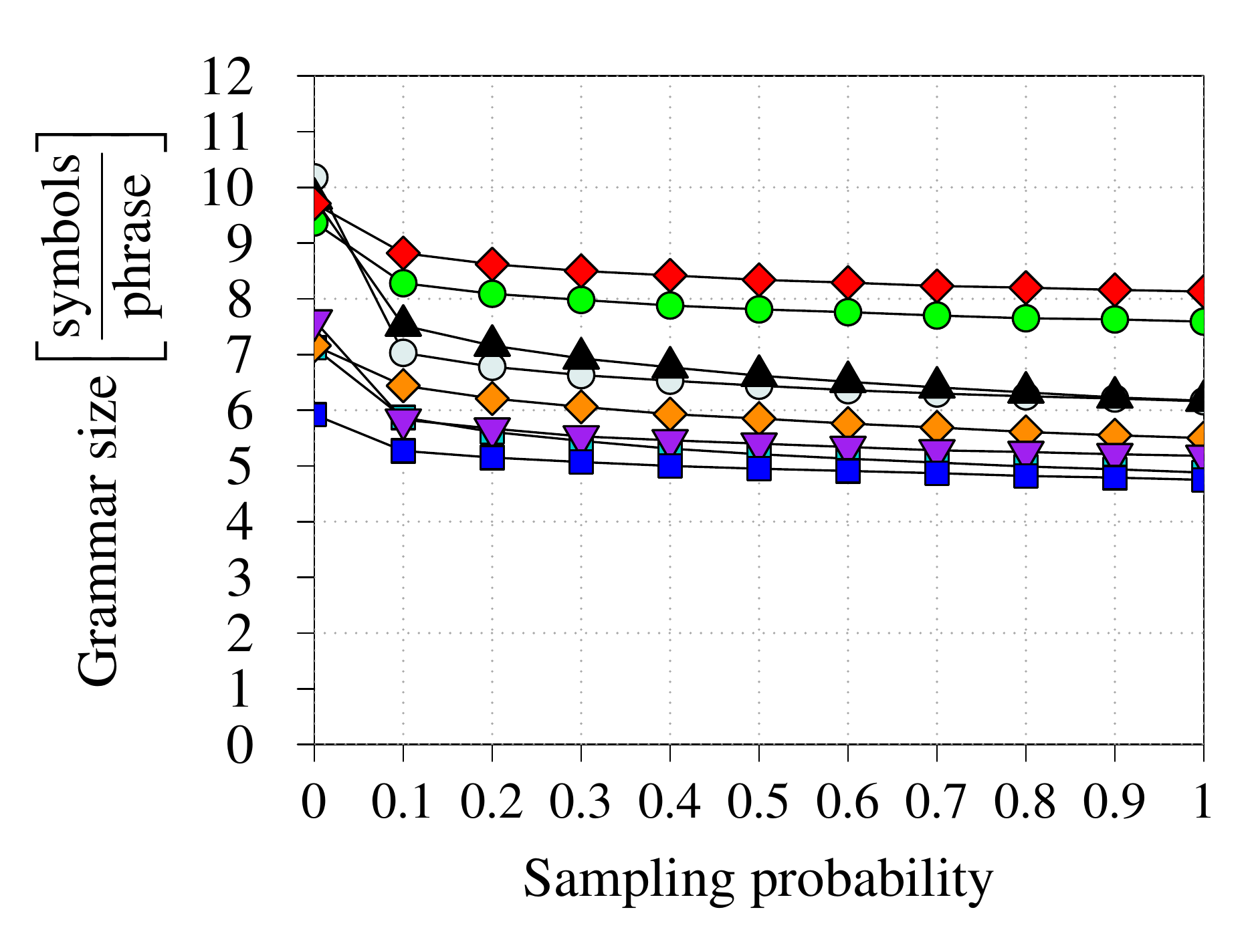}
  \endminipage
  \minipage{0.5\textwidth}
    \includegraphics[trim = 6mm 3mm 0mm 0mm, width=\linewidth]
      {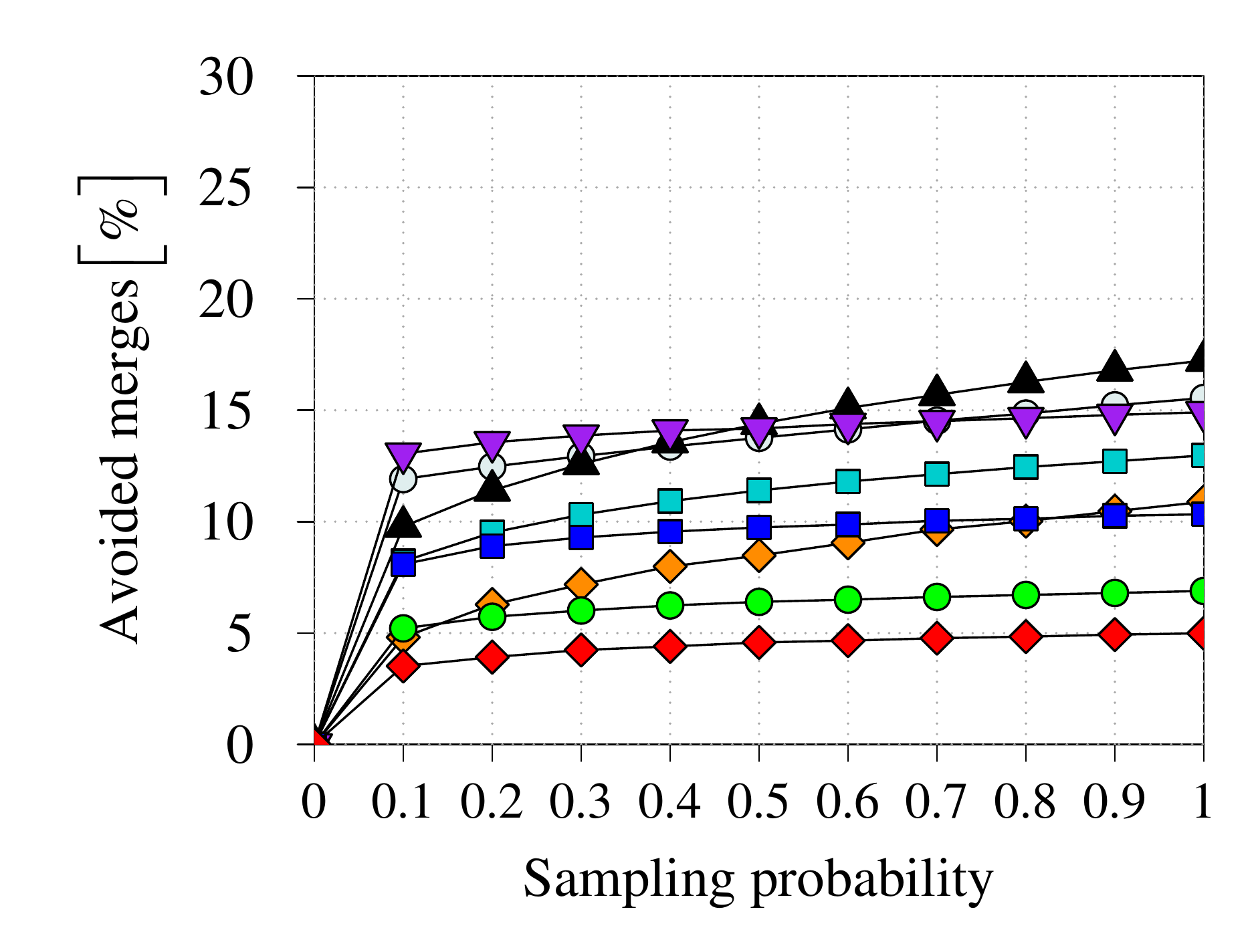}
  \endminipage
  \caption{Performance of the Lazy-AVLG algorithm for different values
    of the parameter $p$ (probability of storing Karp-Rabin
    fingerprint in the hash table) on the files from Pizza \& Chili
    corpus. The graphs in the top row show the normalized RAM usage
    (in bytes per phrase of the LZ77 parsing) and the runtime (in ns
    per symbols of the input text). The bottom row shows the resulting
    grammar size (the total length of right-hand sides of all
    productions) divided by $z$, and the percentage of merges avoided
    during the greedy merge procedure (in \%).}
  \label{fig:sampling}
\end{figure}

The key parameter that controls the runtime, peak RAM usage, and the
size of the grammar in our algorithm is the probability $p$ of
including the Karp-Rabin fingerprint of a nonterminal in the hash
table. In our first experiment, we study the performance of our
Lazy-AVLG implementation for different values of the parameter $p$. We
tested all values $p \in \{0, 0.1, 0.2, \ldots, 1\}$ and for each we
measured the algorithm's time and memory usage, and the size of the
final grammar. The results are given in \cref{fig:sampling}.

While utilizing the Karp-Rabin fingerprints (i.e., setting $p>0$) can
notably reduce the final grammar size (up to 40\% for the cere file),
it is not worth using values $p$ much larger than $0.1$, as it quickly
increases the peak RAM usage (e.g., by about 2.3x for the
sources.001.2 testfile) and this increase is not repaid significantly
in the further grammar reduction. The fourth panel in
\cref{fig:sampling} provides some insight into the reason for this. It
shows the percentage of cases, where during the greedy merging of
nonterminals enclosed by the source of the phrase, the algorithm is
able to avoid merging two nonterminals, and instead use the existing
nonterminal. Having \emph{some} fingerprints in the hash table turns
out to be enough to avoid creating between 4-14\% of the new
nonterminals, but having more does not lead to a significant
difference.  Since peak RAM usage is the likely limiting factor for
this algorithm --- a slower algorithm is still usable, but exhaustion
of RAM can prevent it running entirely --- we chose $p=0.125$ as the
default value in our implementation.  This is the value used in the
next two experiments.

\subparagraph*{Grammar Size.}

\begin{figure}[t!]
  \centering
  \minipage{1.0\textwidth}
    \includegraphics[trim = 0mm 0mm 0mm 0mm, width=\linewidth]
      {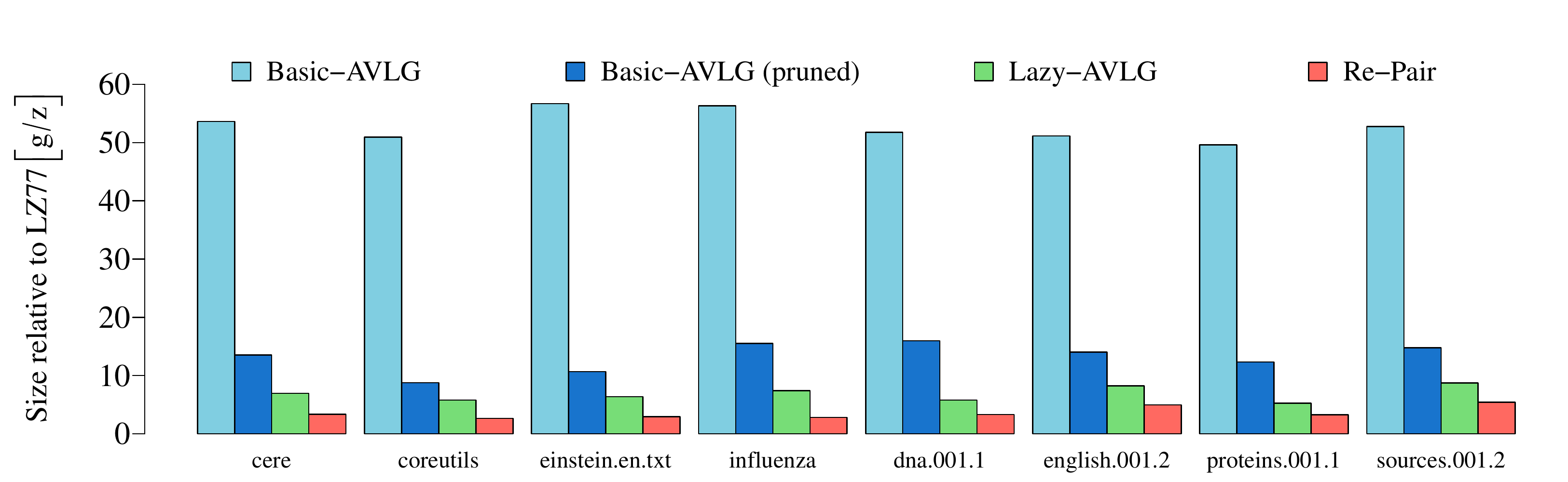}
  \endminipage
  \caption{Comparison of the size (measured as the total length of the
    right hand sides of all nonterminals) of grammars produced by the
    Basic-AVLG, Lazy-AVLG ($p = 0.125$), and Re-Pair algorithms on the
    files from the Pizza \& Chili corpus. Basic-AVLG (pruned) denotes
    the size of grammar produced by Basic-AVLG with all nonterminals
    not reachable from the root removed. All sizes are normalized with
    respect to the size of the LZ77 parsing ($z$).}
  \label{fig:size}
\end{figure}

In our second experiment, we compare the size of the grammar produced
by Lazy-AVLG to Basic-AVLG and Re-Pair.  In the comparison we also
include the size of the grammar obtained by running Basic-AVLG and
removing all nonterminals not reachable from the root. We have run the
experiments on 8/10 testfiles, as running Re-Pair on the large files
is prohibitively time consuming. The results are given in
\cref{fig:size}.

Lazy-AVLG produces grammars that are between 1.59x and 2.64x larger
than Re-Pair (1.95x on average). The resulting grammar is always at
least 5x smaller than produced by Basic-AVLG, and also always smaller
than Basic-AVLG (pruned). Importantly, the RAM usage of our conversion
is proportional the size of the final grammar, whereas the algorithm
to compute the pruned version of Basic-AVLG must first obtain the
initial large grammar, increasing peak RAM usage.  This is a major
practical concern, as described next.

\subparagraph*{Application in the Construction of BWT.}

\begin{figure}[t!]
  \centering
  \hspace{-0.5cm}
  \minipage{0.5\textwidth}
    \includegraphics[trim = 6mm 3mm 0mm 0mm, width=\linewidth]
      {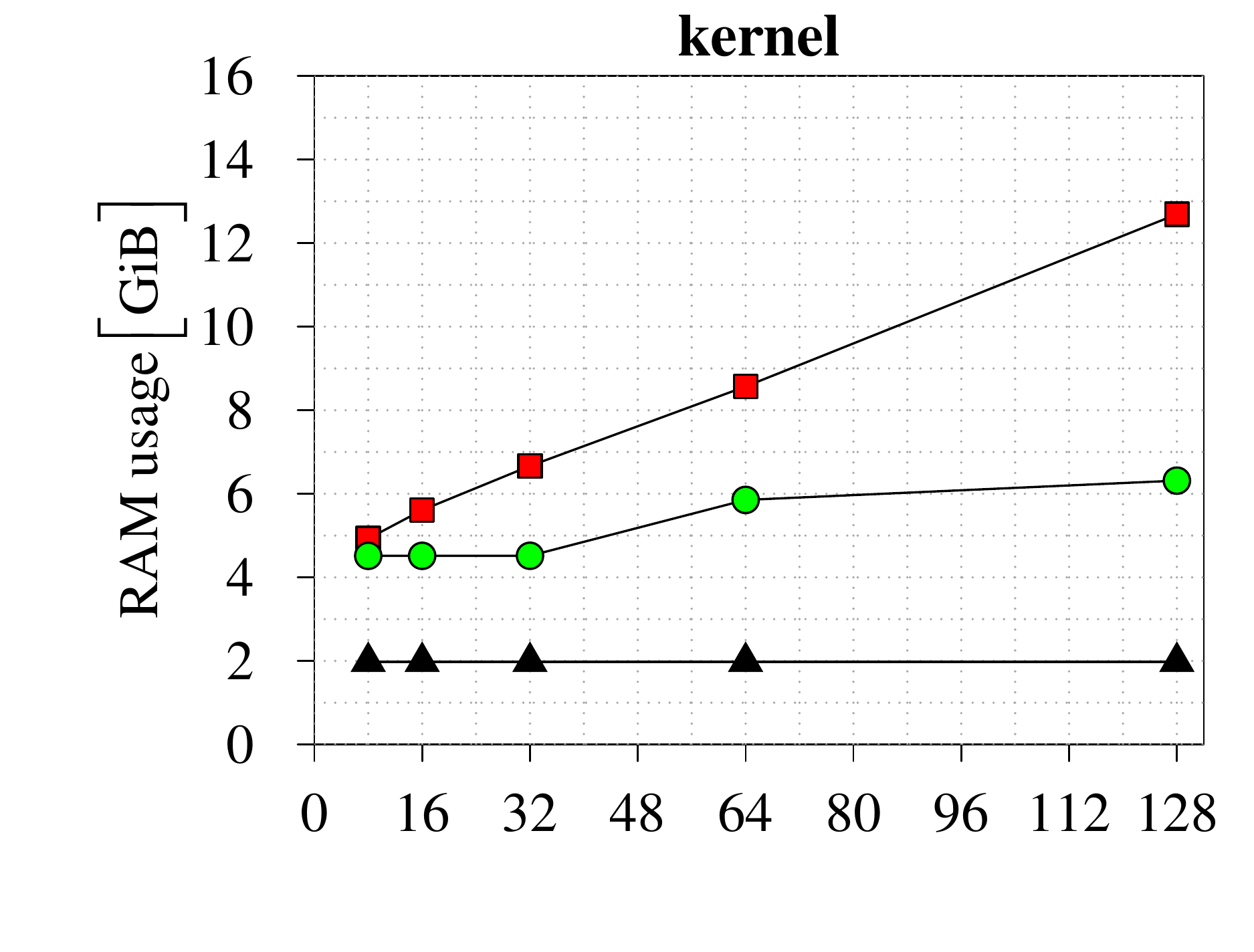}
  \endminipage
  \minipage{0.5\textwidth}
    \includegraphics[trim = 3mm 3mm 3mm 0mm, width=\linewidth]
      {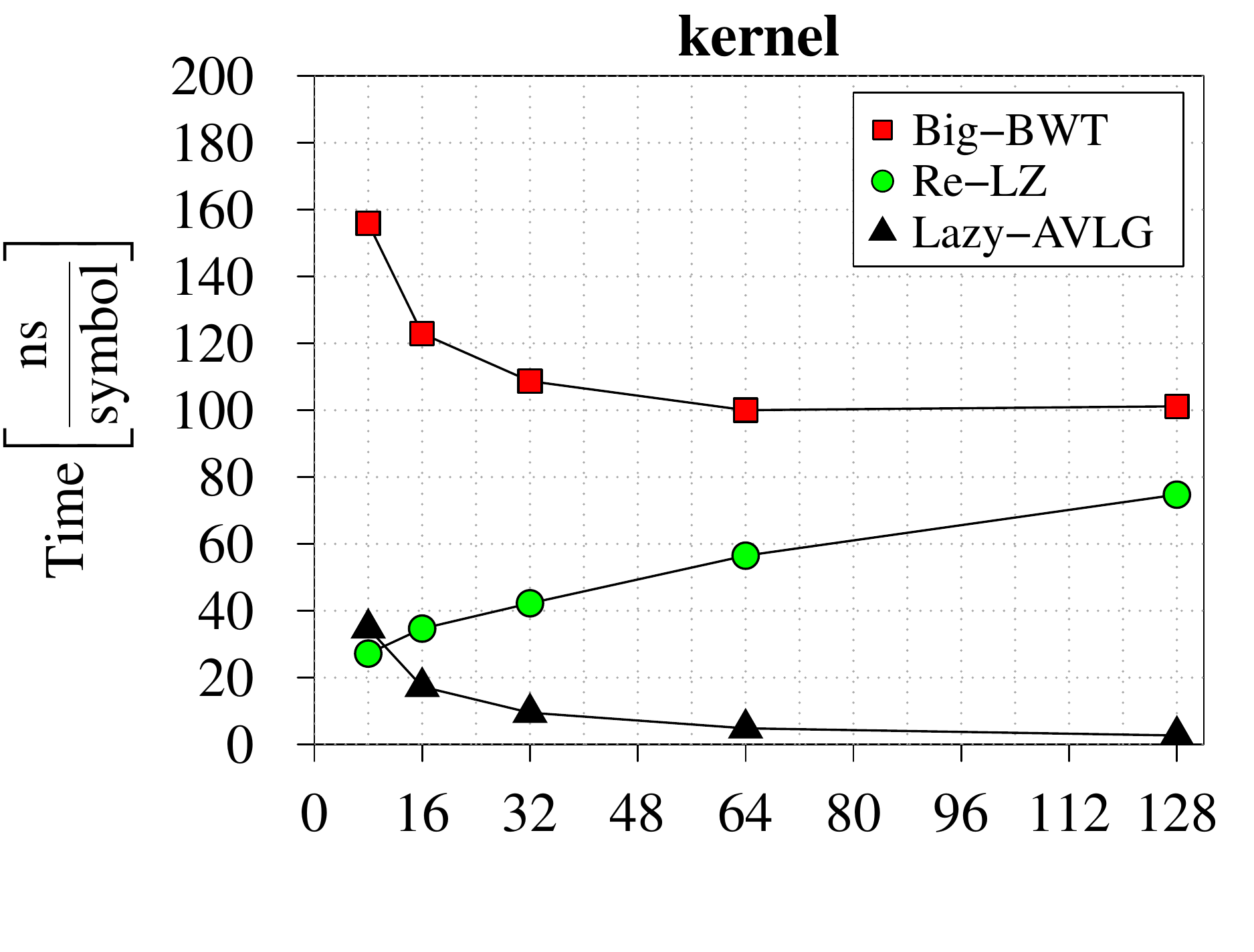}
  \endminipage
  \newline\vspace{-0.4cm}

  \hspace{-0.5cm}
  \minipage{0.5\textwidth}
    \includegraphics[trim = 6mm 3mm 0mm 0mm, width=\linewidth]
      {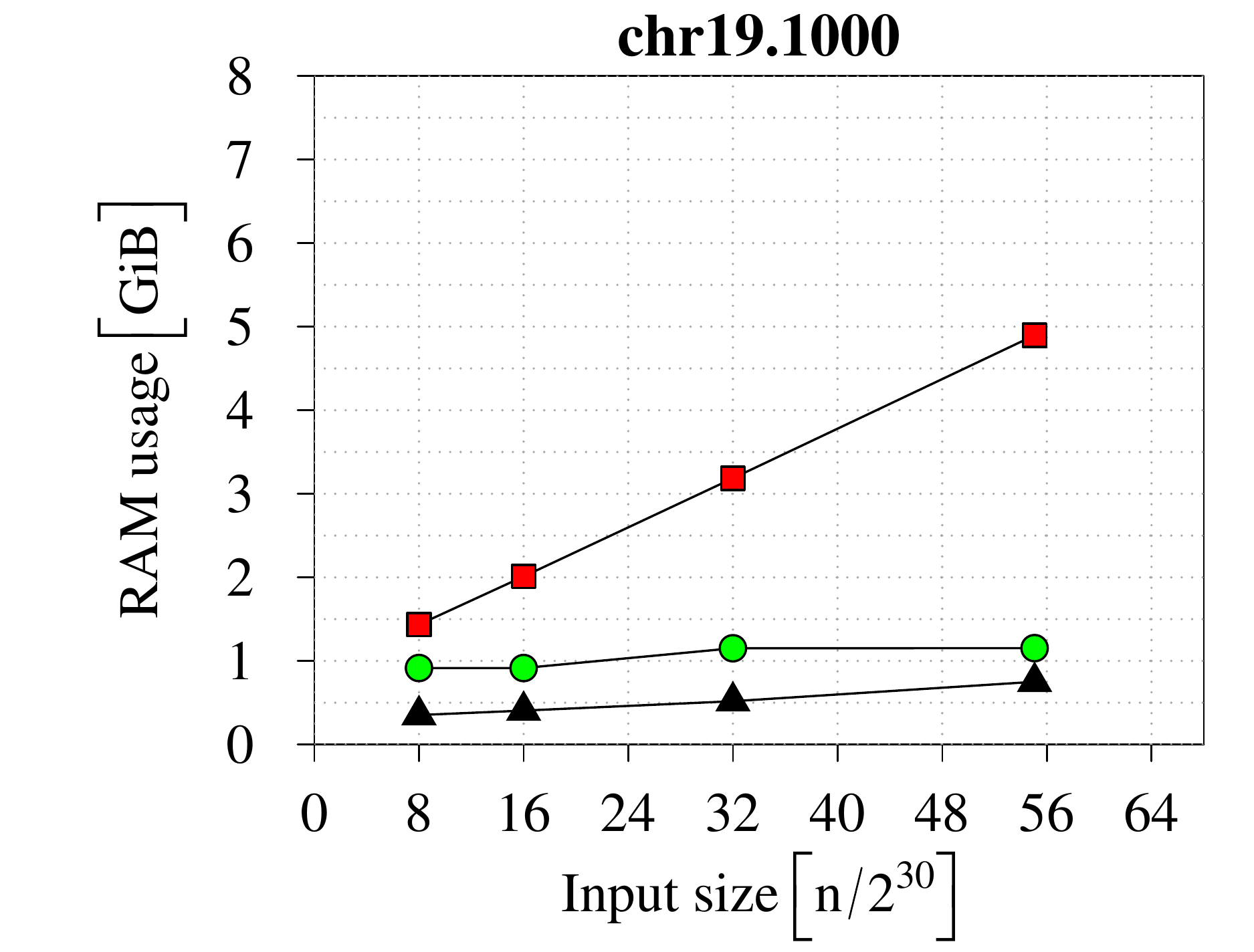}
  \endminipage
  \minipage{0.5\textwidth}
    \includegraphics[trim = 3mm 3mm 3mm 0mm, width=\linewidth]
      {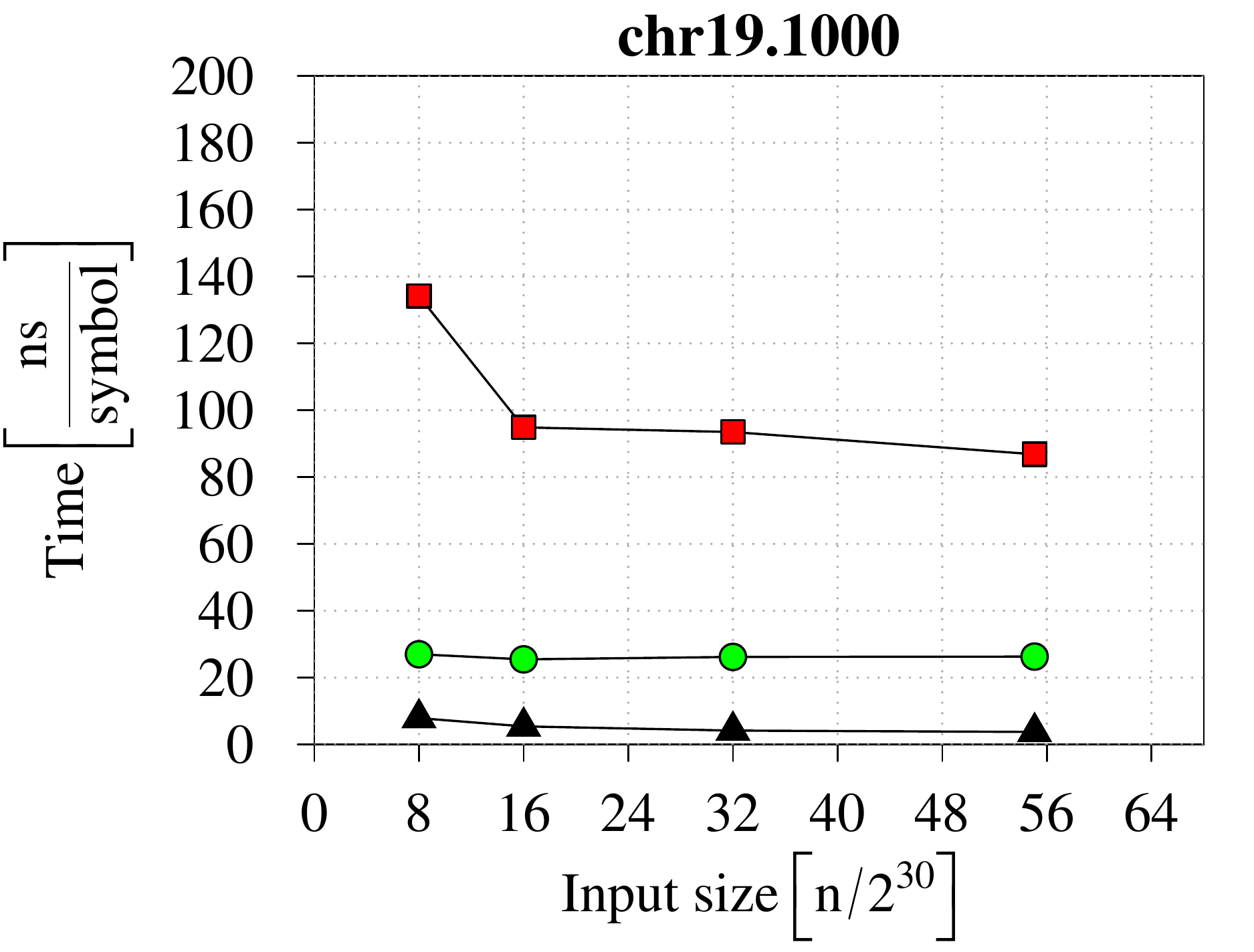}
  \endminipage
  \caption{Scalability of Big-BWT compared to Re-LZ and Lazy-AVLG.
    The graphs on the left show the normalized runtime in ns/char. The
    graphs on the right show the RAM usage in GiB.}
  \label{fig:scalability}
\end{figure}

In our third and main experiment, we evaluate the potential of the
method to construct RLBWT by first computing/approximating the
LZ77-parsing of the text in $\Omega(n)$ time, then converting the
resulting compressed representation of $\T$ into an RLBWT in $\bigO(f
\polylog n)$ time (where $f$ denotes the number of factors) using the
algorithm presented in~\cite{focs2020}. The first step of this
algorithm is the conversion of the input LZ-compressed text into a
straight-line grammar. To our knowledge, the only known approach that
achieves runtime within the $\bigO(f \polylog n)$ time bound is the
algorithm of Rytter~\cite{Rytter03} studied in this paper. Thus, if
the approach of~\cite{focs2020}, has a chance of being practical, the
conversion step needs to be efficient. We remark that this step is a
necessary but not a sufficient condition, and the remaining components
of the algorithm in~\cite{focs2020} need to first be engineered before
the practicality of this approach is fully determined.

We compared the computational performance of Big-BWT, Re-LZ, and our
implementation of Lazy-AVLG with $p = 0.125$ (the default value). We
have evaluated the runtime and peak RAM usage of these three algorithm
on successively longer prefixes of the large testfiles (chr19.1000 and
kernel). The RAM use of Re-LZ was set to match RAM use of Big-BWT on
the shortest input prefix. To allow a comparison with different
methods in the future, we evaluated Lazy-AVLG on the LZ77 parsing
rather than on the output of Re-LZ. Thus, to obtain the performance of
the pipeline Re-LZ + Lazy-AVLG, one should multiply the runtime and
RAM usage of Lazy-AVLG by the approximation ratio $f/z$ of Re-LZ. We
have found the value $f/z$ to not exceed $1.05$ on any of the kernel
prefixes, and $1.27$ on any of the chr19.1000 prefixes (with the peak
reached on the largest prefixes). This puts the RAM use of Lazy-AVLG
on the output of Re-LZ still below that of Re-LZ, and its runtime
still well below other two programs. The results are given in
\cref{fig:scalability}.

At the RAM usage set to stay below that of Big-BWT, the runtime of
Re-LZ is always below that of Big-BWT. The reduction is by a factor of
at least three for all prefixes of chr19.1000, and by at least 25\%
for all prefixes of kernel, with the gap decreasing with the prefix
length. The runtime of Lazy-AVLG stays significantly below that of
both other methods. More importantly, this also holds for Lazy-AVLG's
peak RAM usage.  This shows that construction of the RLBWT via
LZ-parsing has the potential to be the fastest in practice, especially
given that approximation of LZ is the only step requiring $\Omega(n)$
in the pipeline from text to the RLBWT proposed in~\cite{focs2020}. We
further remark that although the construction of RLBWT has received
much attention in recent years, practical approximation of LZ77 is a
relatively unexplored topic, potentially offering significant further
speedup via new techniques or parallelization.  This may be an easier
task since, unlike for any BWT construction algorithm,
LZ-approximation algorithms need not be exact.

\bibliography{paper}

\end{document}